\documentclass[preprint]{aastex}
\usepackage{epsf}
\usepackage{times}

\begin{document}

\title{A complete sample of twelve very X-ray luminous galaxy clusters at $z>0.5$} 

\author{H.\ Ebeling\altaffilmark{1},
E.\ Barrett\altaffilmark{1}, D.\ Donovan\altaffilmark{1}, C.-J.\ Ma\altaffilmark{1}, A.C.\
Edge\altaffilmark{2}, L.\ van Speybroeck\altaffilmark{3}}

\altaffiltext{1}{Institute for Astronomy, University of Hawaii, 2680 Woodlawn
  Drive, Honolulu, HI 96822, USA} 

\altaffiltext{2}{Department of Physics, University of Durham, South Road,
  Durham, DH1 3LE, UK}

\altaffiltext{3}{Harvard-Smithsonian Center for Astrophysics, 60 Garden St,
  Cambridge, MA 02138, USA}

\slugcomment{submitted to ApJ}

\begin{abstract}
We present the statistically complete and cosmologically most relevant subset of
the twelve most distant galaxy clusters detected at $z>0.5$ by the MAssive
Cluster Survey (MACS). Ten of these systems are new discoveries; only two
(MACSJ0018.5+1626 aka CL\,0016+1609, and MACSJ0454.1$-$0300 aka
MS\,0451.6$-$0305) were previously known. We provide fundamental cluster
properties derived from our optical and X-ray follow-up observations as well as
the selection function in tabulated form to facilitate cosmological studies
using this sample.
\end{abstract}


\section{Introduction}

The Massive Cluster Survey (MACS) was launched in 1998 with the goal of
compiling the first large, statistical sample of very X-ray luminous clusters at
$z>0.3$. A comparison of the flux limits and solid angles of various X-ray
cluster surveys conducted previously demonstrates that the ROSAT All-Sky Survey
(RASS) was then -- and remains until today -- the only existing X-ray dataset
that allows these extremely rare systems to be detected in significant numbers
(fig.~1 of Ebeling, Edge \& Henry 2001). A detailed description of the MACS
survey strategy is given in the same paper. In brief, the MACS cluster sample is
compiled by applying limits in X-ray flux and spectral hardness to all sources
listed in the RASS Bright Source Catalogue (BSC, Voges et al.\ 1999) that fall
within the part of the extragalactic sky that is observable from Mauna Kea
($-40^\circ>\delta>80^\circ$, $|b|>20^\circ$) and by conducting an
identification program that combines searching various astronomical object
catalogues with extensive optical imaging observations. Likely cluster
candidates are then targeted in spectroscopic observations and added to the MACS
catalogue if their measured redshift exceeds $z=0.3$. Although the required
optical follow-up observations are currently complete only for the MACS Bright
Sample (defined as comprising all sources with BSC detect fluxes in excess of
$2\times 10^{-12}$ erg$^{-1}$ cm$^{-2}$ in the 0.1--2.4 keV band, Ebeling et
al.\ 2007, in preparation), we gave priority to the most distant cluster
candidates regardless of X-ray flux down to the overall MACS flux limit of
$1\times 10^{-12}$ erg$^{-1}$ cm$^{-2}$. We are thus confident that the sample
presented here, the set of the twelve MACS clusters at $z>0.5$, is complete at
the nominal 90\% level aimed at for the entire survey.

We assume a $\Lambda$CDM cosmology with $\Omega_m=0.3$, $\Omega_\Lambda=0.7$ and
$H_o=70$\,km\,s$^{-1}$\,Mpc$^{-1}$ throughout.

\section{Follow-up observations}

As they were detected in the course of the survey, MACS clusters at $z>0.5$ were
given highest priority in our in-depth follow-up observations. Eight of the
twelve clusters forming the final sample were targeted in dedicated observations
at 30 GHz with the BIMA interferometer; a strong detection of the
Sunyaev-Zeldovich effect was obtained for all of them (LaRoque et
al.\ 2003). All newly discovered clusters were observed with the ACIS-I
instrument aboard the Chandra X-ray Observatory in guaranteed time made
available by Chandra Telescope Scientist Leon van Spectroscopy. In December 2006
we completed moderately deep observations with the SuprimeCam wide-field imager
on the Subaru 8.3m telescope on Mauna Kea in the B, V, R, I, and z$\prime$
passbands. In addition, U-band images of all systems have been obtained with the
MegaCam 1-degree imager on the 3.6m Canada-France-Hawaii
Telescope. High-resolution imaging of the cluster cores was performed with the
Advanced Camera for Surveys on the Hubble Space Telescope (Program ID 09722, PI
Ebeling) in the F555W and F814W filters. Returning to groundbased observations,
we conducted extensive spectroscopic observations of galaxies in the fields of
all twelve clusters with multi-object spectrographs on 8m-class telescopes on
Mauna Kea (Gemini, Keck-I, Keck-II); a catalogue of more than 3,000 redshifts of
galaxies in MACS cluster fields can be found in Barrett \& Ebeling (2007).
 
A wealth of science results, ranging from an in-depth study of the massive
cooling-core cluster MACSJ1423.8$+$2404, through work on galaxy evolution in
clusters, to a study of the large-scale structure surrounding the most distant
MACS clusters, has been obtained from these observations (LaRoque et al.\ 2003,
Ebeling et al.\ 2004, Allen et al.\ 2004, Stott et al.\ 2007, Smith et
al.\ 2007, Barrett et al.\ 2007, Kartaltepe et al.\ 2007, Ebeling et al.\ 2007,
Edge et al.\ 2007) with additional major projects such as a comprehensive
weak-lensing analysis of the cluster mass distribution and two complementary
studies yielding strong cosmological constraints on dark matter and dark energy
using MACS being well advanced (Donovan et al., in preparation; Mantz et al., in
preparation; Allen et al., in preparation).

\section{The twelve most distant MACS clusters}

To facilitate the use of this statistically complete sample in studies conducted
by the scientific community in general we here present the MACS $z>0.5$
catalogue listing the clusters' fundamental physical properties, such as X-ray
position, redshift, radial velocity dispersion, X-ray fluxes and luminosities,
and X-ray gas temperatures (Table~1). To allow the computation of search volumes
for cosmological and astrophysical applications we also list the X-ray detect
flux from the RASS BSC and provide a look-up table of the MACS selection
function in Table~2. 

Color images of the central $10\times 10$ arcmin$^2$ of each cluster field
(corresponding to roughly 3.8 Mpc at the median redshift of the clusters in our
sample) are shown in Fig.~1. All images are based on VRz$\prime$ imaging
obtained with SuprimeCam in sub-arcsec seeing. Overlaid are isodensity contours
of the adaptively smoothed X-ray emission in the $0.5-7$ keV band as observed
with Chandra. Note the wide range of X-ray morphologies, indicating that the
MACS sample is neither biased in favor of cooling-core systems (which feature
the highest central X-ray surface brightness levels) nor in favor of heavily
disturbed clusters (which may be temporarily boosted in their X-ray luminosity
and temperature due to merger-induced shocks). A qualitative assessment of the
morphology of the cluster gas as traced by the X-ray surface brightness
distribution is given in the final column of Table~1.

\section{Summary}

We present the statistically complete sample of the twelve most distant MACS
clusters, all of which feature redshifts of $z>0.5$ and ten of which are new
discoveries. Extensive follow-up observations at wavelengths from the X-ray to
the radio passbands have confirmed that these are indeed exceptionally massive
systems and thus the high-redshift counterparts of the best studied massive
clusters in the local universe. Results from several studies on individual
clusters or the sample as a whole have been published; additional studies,
ranging from a comprehensive weak-lensing analysis to cosmological applications
of the MACS sample, are well advanced. In recognition of the legacy character of
this sample we provide in this paper an overview of fundamental properties of
all twelve clusters as well as a tabulated version of the MACS selection
function to facilitate the use of our sample by the scientific community for a
wide range of astrophysical and cosmological applications.

\vspace*{1cm}\mbox{}
\noindent
We dedicate this paper to the memory of our friend and colleague Leon van
Speybroeck whose advice and enthusiasm (not to mention his Guaranteed Time on
Chandra) inspired all of us. We miss you, Leon. HE acknowledges financial
support from NASA and SAO (grants NAG 5-8253 and GO2-3168X, respectively).

\references
Allen S.W., Schmidt R.W, Ebeling H., Fabian A.C., van Speybroeck L., 2004, MNRAS, 353, 457\\
Barrett E.\ \& Ebeling H. 2007, MNRAS, submitted\\
Barrett E., Ebeling H.\ \& Ma C.-J., 2007, MNRAS, submitted\\
Ebeling H., Edge A.C.\ \& Henry J.P. 2001, ApJ, 553, 668\\
Ebeling H., Barrett E.\ \& Donovan D., 2004, ApJ, 609, 49L\\
Ebeling H., White D.A.\ \& Rangarajan F.V.N. 2006, MNRAS, 368, 65\\
Ebeling H.\ et al. 2007, MNRAS, submitted\\
Edge A.C.\ et al. 2007, MNRAS, in press\\
Kartaltepe J., Ebeling H., Donovan D., Ma C.-J. 2007, MNRAS, submitted\\
LaRoque S.\ et al.\ 2003, ApJ, 583, 559\\
Smith G.\ et al. 2007, MNRAS, submitted\\
Stott J.\ et al. 2007, ApJ, in press\\
Voges, W.\ et al. 1999, A\&AS, 349, 389

\begin{deluxetable}{ccccccrrrrrc}
\tabletypesize{\scriptsize}
\tablecaption{Fundamental cluster properties for the MACS $z>0.5$ sample. \label{props}}
\tablewidth{0pt}
\tablehead{
\colhead{MACS name} & \colhead{$\alpha$ (J2000)} & \colhead{$\delta$ (J2000)} & \colhead{z} & \colhead{$n_z$} & \colhead{$\sigma$ (km/s)} & \colhead{$f_{\rm X,det,BSC}$} & \colhead{$L_{\rm X,BSC}$} & \colhead{$f_{\rm X,Chandra}$} & \colhead{$L_{\rm X,Chandra}$} & \colhead{k$T$ (keV)} &\colhead{morph.\ code} 
}
\startdata
MACSJ0018.5$+$1626  & 00 18 33.835 &  $+$16 26 16.64 & 0.5461  &   51 & 1420                    &   $1.33\pm 0.36$  & $14.2\pm 3.9$ & $2.14\pm 0.03$& $19.6\pm 0.3$ & $9.4\pm 1.3$ & 3\\
MACSJ0025.4$-$1222  & 00 25 29.381 &  $-$12 22 37.06 & 0.5843  &   31 & \mbox{\hspace{1mm}} 730 &   $1.10\pm 0.26$  & $13.7\pm 3.3$ & $0.81\pm 0.02$& $ 8.8\pm 0.2$ & $7.1\pm 0.7$ & 3\\
MACSJ0257.1$-$2325  & 02 57 09.151 &  $-$23 26 05.83 & 0.5053  &   24 & \mbox{\hspace{1mm}} 900 &   $1.23\pm 0.28$  & $11.3\pm 2.6$ & $1.80\pm 0.03$& $13.7\pm 0.3$ &$10.5\pm 1.0$ & 2\\
MACSJ0454.1$-$0300  & 04 54 11.125 &  $-$03 00 53.77 & 0.5377  &   27 & 1250                    &   $1.55\pm 0.32$  & $17.1\pm 3.5$ & $1.88\pm 0.04$& $16.8\pm 0.6$ & $7.5\pm 1.0$ & 2\\
MACSJ0647.7$+$7015  & 06 47 50.469 &  $+$70 14 54.95 & 0.5908  &   37 & \mbox{\hspace{1mm}} 900 &   $1.34\pm 0.29$  & $16.8\pm 3.6$ & $1.49\pm 0.03$& $15.9\pm 0.4$ &$11.5\pm 1.0$ & 2\\
MACSJ0717.5$+$3745  & 07 17 30.927 &  $+$37 45 29.74 & 0.5460  &  139\mbox{\hspace{1mm}} & 1640 &   $1.92\pm 0.32$  & $20.9\pm 3.5$ & $2.74\pm 0.03$& $24.6\pm 0.3$ &$11.6\pm 0.5$ & 4\\
MACSJ0744.8$+$3927  & 07 44 52.470 &  $+$39 27 27.34 & 0.6972  &   40 & 1050                    &   $1.19\pm 0.26$  & $21.0\pm 4.6$ & $1.44\pm 0.03$& $22.9\pm 0.6$ & $8.1\pm 0.6$ & 2\\
MACSJ0911.2$+$1746  & 09 11 11.277 &  $+$17 46 31.94 & 0.5049  &   26 & 1150                    &   $1.05\pm 0.26$  &  $9.8\pm 2.5$ & $1.00\pm 0.02$& $ 7.8\pm 0.3$ & $8.8\pm 0.7$ & 4\\
MACSJ1149.5$+$2223  & 11 49 35.093 &  $+$22 24 10.94 & 0.5445  &   64 & 1810                    &   $1.20\pm 0.24$  & $13.0\pm 2.6$ & $1.95\pm 0.04$& $17.6\pm 0.4$ & $9.1\pm 0.7$ & 4\\
MACSJ1423.8$+$2404  & 14 23 47.663 &  $+$24 04 40.14 & 0.5428  &   45 & 1310                    &   $1.03\pm 0.23$  & $11.3\pm 2.5$ & $1.80\pm 0.06$& $16.5\pm 0.7$ & $7.0\pm 0.8$ & 1\\
MACSJ2129.4$-$0741  & 21 29 26.214 &  $-$07 41 26.22 & 0.5891  &   47 & 1460                    &   $1.05\pm 0.28$  & $12.6\pm 3.3$ & $1.45\pm 0.03$& $15.7\pm 0.4$ & $8.1\pm 0.7$ & 3\\
MACSJ2214.9$-$1359  & 22 14 57.415 &  $-$14 00 10.78 & 0.5026  &   65 & 1290                    &   $1.42\pm 0.33$  & $12.6\pm 2.9$ & $1.85\pm 0.03$& $14.1\pm 0.3$ & $8.8\pm 0.7$ & 2\\
\enddata
\tablecomments{The listed X-ray centroids are determined from Chandra ACIS-I
  data; redshifts, velocity dispersions and number of redshifts are determined
  within a circle of 1 Mpc radius from the X-ray centroid. X-ray fluxes ({\em
    detect fluxes} in units of $10^{-12}$ erg s$^{-1}$ cm$^{-2}$) and
  luminosities ({\em total luminosities} in units of $10^{44}$ erg s$^{-1}$) are
  for the 0.1$-$2.4 keV band. MACSJ0018.5$+$1626 and MACSJ0454.1$-$0300 are
  rediscoveries of CL\,0016+1609 and MS\,0451.6$-$0305, respectively. Fluxes and
  luminosities from Chandra exclude X-ray point sources and have been
  extrapolated to a radius of $r_{\rm 200}$. X-ray temperatures were measured
  from Chandra data within $r{\rm 1000}$ but excluding a central region of 70
  kpc radius around the listed X-ray centroid. Morphology is assessed visually
  based on the amount of substructure apparent in the X-ray contours as well as
  overall elongation or sphericity -- the assigned codes range from 1
  (apparently relaxed) to 4 (extremely disturbed). }
\end{deluxetable}

\begin{deluxetable}{rrrr}
\tabletypesize{\scriptsize}
\tablecaption{MACS selection function. \label{props}}
\tablewidth{0pt}
\tablehead{
\colhead{$f_{\rm det, BSC}$} & \colhead{solid angle} & \colhead{$f_{\rm det, BSC}$} & \colhead{solid angle} 
}
\startdata
  0.550 &      0 &     0.925 &  10136 \\ 
  0.575 &     20 &     0.950 &  11328 \\ 
  0.600 &     94 &     0.975 &  12501 \\ 
  0.625 &    251 &     1.000 &  13544 \\ 
  0.650 &    667 &     1.100 &  16257 \\ 
  0.675 &   1253 &     1.200 &  17823 \\ 
  0.700 &   2071 &     1.300 &  18909 \\ 
  0.725 &   2877 &     1.400 &  19612 \\ 
  0.750 &   3882 &     1.500 &  20043 \\ 
  0.775 &   4708 &     2.000 &  21123 \\ 
  0.800 &   5464 &     3.000 &  21886 \\ 
  0.825 &   6205 &     5.000 &  22297 \\ 
  0.850 &   7020 &    10.000 &  22488 \\ 
  0.875 &   7940 &    50.000 &  22617 \\ 
  0.900 &   8999 & & \\
\enddata
\tablecomments{RASS detect fluxes $f_{\rm det, BSC}$ in units of $10^{-12}$ erg
  s$^{-1}$ cm$^{-2}$ in the 0.1$-$2.4 keV band and solid angle in square degrees
  covered at fluxes exceeding $f_{\rm det, BSC}$.}
\end{deluxetable}

\clearpage

\begin{figure}
\vspace*{-9mm}
\caption{VRz$\prime$ color images of the twelve MACS clusters at $z>0.5$ as obtained
  with the Subaru SuprimeCam wide-field imager. Exposure times range from 20 to
  45 minutes in the three passbands. Only the central $10\times 10$ arcmin$^2$
  region is shown. Overlaid are logarithmically spaced isodensity contours of
  the adaptively smoothed X-ray surface brightness in the 0.5$-$7 keV band as
  observed with Chandra's ACIS-I detector. We use the {\sc Asmooth} algorithm of
  Ebeling, White \& Rangarajan (2006) and require a minimal significance of
  $3\sigma$ for all features in the adaptively smoothed X-ray image. (files too large for astroph -- all images available from first author upon request)}
\end{figure}

\end{document}